\documentclass[conference]{IEEEtran}

\usepackage{graphicx} 
\usepackage[table]{xcolor}
\usepackage{xcolor}
\usepackage{listings}
\usepackage{tabularx}
\usepackage{tcolorbox}
\usepackage{todonotes}[]
\usepackage{booktabs}
\usepackage{tikz}
\usepackage{tcolorbox}
\usepackage{fbox}
\usepackage{multirow}
\usepackage[flushleft]{threeparttable}
\usepackage{xurl}
\usepackage[hidelinks]{hyperref}
\usepackage{float}
\usepackage[sort]{cite}
\usepackage{balance}

\usepackage{microtype} 

\usepackage{algorithm}

\usepackage{algpseudocode}

\usepackage[flushleft]{threeparttable}

\definecolor{lightgray}{rgb}{.9,.9,.9}
\definecolor{darkgray}{rgb}{.4,.4,.4}
\definecolor{purple}{rgb}{0.65, 0.12, 0.82}
\definecolor{light-gray}{gray}{0.80}

\lstdefinelanguage{JavaScript}{
  keywords={typeof, new, true, false, catch, function, return, null, catch, switch, var, if, in, while, do, else, case, break, const},
  keywordstyle=\color{blue}\bfseries,
  ndkeywords={class, export, boolean, throw, implements, import, this, it, expect, test, require},
  ndkeywordstyle=\color{darkgray}\bfseries,
  identifierstyle=\color{black},
  sensitive=false,
  comment=[l]{//},
  morecomment=[s]{/*}{*/},
  commentstyle=\color{purple}\ttfamily,
  stringstyle=\color{red}\ttfamily,
  morestring=[b]',
  morestring=[b]"
}

\newcommand{\Space}[1]{}
\newcommand{\Fix}[1]{\textcolor{black}{#1}}

\newcommand{\odtest}{order-dependent test}
\newcommand{\victim}{victim}
\newcommand{\brittle}{brittle}
\newcommand{\polluter}{polluter}
\newcommand{\statesetter}{state-setter}

\newcommand{\describe}{\texttt{describe}}
\newcommand{\testFunc}{\texttt{test}}
\newcommand{\itFunc}{\texttt{it}}
\newcommand{\block}{block}
\newcommand{\Block}{Block}
\newcommand{\reorder}{reorder}
\newcommand{\project}{project}
\newcommand{\Project}{Project}
\newcommand{\jsTod}{\texttt{JS-TOD}}

\newcommand{\NumReorders}{10}
\newcommand{\NumReruns}{10}
\newcommand{\NumInitialProjects}{700}
\newcommand{\NumJestProjects}{304}
\newcommand{\NumProjectsWithWorkingTests}{82}
\newcommand{\NumEvaluatedProjects}{81}
\newcommand{\NumProjectsWithSingleTestInDescribe}{12}
\newcommand{\NumProjectsToExclude}{14}
\newcommand{\NumFinalProjectsForAnalysis}{67}

\newcommand{\NumODTests}{55}
\newcommand{\NumODTestsMocks}{39}

\newcommand{\NumODTestsFiles}{13}
\newcommand{\NumODTestsMocksAll}{42}
\newcommand{\NumProjectsWithODTests}{10}
\newcommand{\NumODTestsAtTestLevel}{52}
\newcommand{\NumODTestsAtDescribeLevel}{three}

\title{Detecting and Evaluating Order-Dependent Flaky Tests in JavaScript}

\author{
    \IEEEauthorblockN{Negar Hashemi\IEEEauthorrefmark{1}, Amjed Tahir\IEEEauthorrefmark{1}, Shawn Rasheed\IEEEauthorrefmark{2}, August Shi\IEEEauthorrefmark{3}, Rachel  Blagojevic\IEEEauthorrefmark{1}}
    \IEEEauthorblockA{\IEEEauthorrefmark{1}Massey University, New Zealand
    \\\{n.hashemi;a.tahir;r.v.blagojevic\}@massey.ac.nz}
    \IEEEauthorblockA{\IEEEauthorrefmark{2}Universal College of Learning, New Zealand
    \\
    s.rasheed@ucol.co.nz}

    \IEEEauthorblockA{\IEEEauthorrefmark{3} The University of Texas at Austin, USA
    \\
    august@utexas.edu}
}

\begin{document}

\maketitle

\begin{abstract}
Flaky tests pose a significant issue for software testing. A test with a non-deterministic outcome may undermine the reliability of the testing process, making tests untrustworthy. 
Previous research has identified 
test order dependency 
as one of the most prevalent causes of flakiness, particularly in Java and Python. However, little is known about test order dependency in JavaScript tests. 
This paper aims to investigate test order dependency in JavaScript projects that use Jest, 
a widely used JavaScript testing framework. We implemented a systematic approach to randomise tests, test suites and \describe{} \block{}s and produced 10 unique test reorders for each level. 
We reran each order 
10 times (100 reruns for each test suite/project) and recorded any changes in test outcomes. We then manually analysed each case that showed flaky outcomes to determine the cause of flakiness. We examined our detection approach on a dataset of \NumEvaluatedProjects{} projects obtained from GitHub.
Our results revealed \NumODTests{} \odtest{}s across \NumProjectsWithODTests{} projects. Most \odtest{}s (\NumODTestsAtTestLevel{}) occurred between tests, while the remaining \NumODTestsAtDescribeLevel{} occurred between \describe{} \block{}s. Those \odtest{}s are caused by either \textit{shared files} (\NumODTestsFiles{}) or \textit{shared mocking state} (\NumODTestsMocksAll{}) between tests. While sharing files is a known cause of \odtest{}s in other languages, our results underline a new cause (shared mocking state) that was not reported previously.
 
\end{abstract}

\begin{IEEEkeywords}
Flaky Tests, Test Order Dependency, JavaScript, Jest
\end{IEEEkeywords}

\section{Introduction}
\label{sec:intro}
Flaky tests (i.e., tests with non-deterministic outcomes) may pass on some runs and fail on others on the same version of the code.
Flaky tests are known to negatively impact product quality and delivery~\cite{fowler2011eradicating,SandhuTesting2015,palmer2019}, as well as testing-related activities~\cite{machalica2019predictive,lam2020dependent}.
A flaky test that fails may not indicate any bug in the code, and developers waste time trying to identify non-existent bugs.
Flaky tests can then lead developers to lose trust in their builds and ignore test failures~\cite{thorve2018empirical},
resulting in additional costs for software projects. 
Indeed, flaky tests present a significant challenge in the software industry~\cite{googleFlaky2016,fowler2011eradicating,palmer2019,Machalica2020Probabilistic}.
A report from Microsoft noted that flaky tests could cost up to \$7.2 million per year for Microsoft Dynamics product line~\cite{herzig2015art}.
GitHub also reported that 9\% of their commits had at least one failed build caused by flaky tests~\cite{Reducing45:online}.

Previous studies have identified and categorized potential causes of flaky tests across different programming languages, including Java~\cite{luo2014empirical}, Python~\cite{gruber2021empirical}, and JavaScript~\cite{Hashemi2022flakyJS}.
Luo et al.~\cite{luo2014empirical} reported the first extensive empirical study on flaky tests in open-source software, and they categorized the causes and fixing strategies for these flaky tests.
They found that one of the major causes of flaky tests is test order dependency—that is, the tests' outcomes depend on the order in which they are executed.
Flaky tests that pass in one order but fail in another are known as \emph{\odtest{}s}.
While test order dependency is acknowledged as one of the top causes of test flakiness in several languages, including Java and Python~\cite{luo2014empirical,thorve2018empirical,eck2019understanding,gruber2021empirical}, the number of \odtest{}s reported in JavaScript projects is low~\cite{Hashemi2022flakyJS}.
However, these findings were based on the reported flaky tests in open-source projects (i.e., extracted from issue trackers). 
The projects may contain \odtest{}s that the developers have yet to discover, potentially failing and misleading developers later during development.
Indeed, the most popular JavaScript testing framework, Jest~\cite{yost2023finding,taleb2023frameworks}, introduced a feature that allows developers to randomize the order in which they run Jest tests due to concerns from developers about the potential of having \odtest{}s~\cite{jest_issue_4386}.
This feature is still relatively new (made available in late 2023) and not widely adopted, meaning many projects may still contain \odtest{}s that developers are unaware of.

In this study, we conduct an in-depth investigation into \odtest{}s in JavaScript, focusing on the Jest testing framework.
Jest offers many features, such as parallelization and ordered test execution, but tests may still
share state between each other, leading to \odtest{}s.
Our goal is to understand the prevalence of \odtest{}s within popular open-source JavaScript projects after actively searching for them. We also want to understand the reasons for their failures, i.e., what state these tests share between them. The results of our study can provide better insights towards how developers can manage \odtest{}s in their test suites, debugging their root causes to eventually fix them so they no longer have different test outcomes in different orders.

The main contributions of this paper are the answers to the following two research questions:




 \begin{itemize}
    \item \noindent \textbf{RQ1} How can \odtest{}s manifest their failures in JavaScript projects?
    \item \noindent \textbf{RQ2} What are the main causes of \odtest{}s?

 \end{itemize}

RQ1 aims to investigate ways that \odtest{}s' failures can manifest in Jest.
We employed a multi-layer reorder and rerun approach to uncover \odtest{}s in Jest.
By altering the sequence in which the tests are executed within programs/test suites and rerunning them multiple times, we observe whether the test outcomes change.
RQ2 aims to understand the main causes of \odtest{}s in Jest.
We manually inspected each detected \odtest{} and classified the cause of flakiness into different categories.

To answer our research questions, we developed a tool called \jsTod, which extracts, reorders, and reruns tests and test
suites using user-defined numbers of reorders and reruns. We evaluated on a dataset of \NumEvaluatedProjects{} projects. Using \jsTod, we reordered tests and test suites 10 times each and rerun them 10 times each. Our results show that \odtest{}s are possible despite Jest executing tests sequentially (which is supposed to guarantee the default test order).

Of the \NumODTests{} detected \odtest{}s, \NumODTestsAtTestLevel{} were found between individual tests, and three were detected between describe blocks within the same test suite rather than between different test suites. 
We found that the main causes of those flaky tests are shared files (e.g., access to a shared file created in one of the tests and accessed by other tests) and shared mocking state (e.g., some mocked state was changed by tests).


\section{Background}
\label{sec:background}
	
\subsection{Test Order-Dependency}

Ideally, tests should function independently of each other and produce the same results regardless of their order of execution~\cite{zhang2014empirically}.
In reality, tests often have different outcomes when executed in different orders~\cite{lam2020dependent,zhang2014empirically}.
These tests are called \emph{\odtest{}s}, and
previous studies found them to form a significant proportion of flaky tests~\cite{luo2014empirical,gruber2021empirical}.

Prior work classified \odtest{}s into one of two categories based on their result when run in isolation: \victim{} or \brittle{}~\cite{Shi2019iFixFlakies}.
A \emph{\victim{}} passes when run in isolation but fails when run after another test(s), known as a \emph{\polluter{}}.
Essentially, a \polluter{} ``pollutes'' the state shared between the two tests, leading the \victim{} to run in an improperly initialized state, resulting in a failure.
On the other hand, a \emph{\brittle{}} test fails when run in isolation but passes when run after another test(s), called a \emph{\statesetter{}}.
Essentially, the \statesetter{} is setting up the initial state the \brittle{} needs to start in for it to pass; otherwise, the \brittle{} fails as it cannot run properly on its own.

Listing~\ref{lst:exp} shows an example of an \odtest{} from ``react-testing-library''~\cite{reactTests}. 
Test ``first'' is a \statesetter{} for test ``second,'' which is a \brittle{} test. Running the test ``second'' before the test ``first'' or in isolation will result in a test failure as the document does not have the proper body to be checked.

\begin{lstlisting}[caption={Order-dependent tests from "react-testing-library"},label={lst:exp}]
test('first', () => {
  render(<div>hi</div>)})
test('second', () => {
 expect(document.body.innerHTML).toEqual('<div><div>hi</div></div>')})
 \end{lstlisting}


\subsection{Jest}
\label{sec:jest}
Jest is the most popular JavaScript testing framework~\cite{taleb2023frameworks,stateofjs2023}\Space{ due to its ease of use and minimal configuration requirements}.
In Jest, a \emph{test suite} corresponds to a JavaScript file that contains the tests to be run, so there can be many different test suites in a project.
A test suite may encompass multiple testing blocks.
A developer may choose to use \describe{} \block{}s to group together multiple related tests~\cite{jestDescribeApi}, introducing a level of organization for their tests.
The \testFunc{} and \itFunc{} functions 
define individual tests within the test suites, each tasked with verifying a specific scenario. 
The \describe{} \block{} allows for hierarchical organization, while the \testFunc{} function outlines the specifics of each case, contributing to a formalized and systematic testing approach.
Note that individual tests do not always need to be contained with a \describe{} \block{}, but if there are \describe{} \block{}s, then tests within the same \describe{} \block{} are executed with each other without interleaving with tests in other \describe{} \block{}s.

Jest has different approaches to running tests and test suites.
By default, Jest runs test suites in parallel rather than sequentially.
By doing so, Jest does not guarantee the running order of test suites\cite{runningOrder}. 
Jest assigns a worker to run each test suite.
A developer can use the option \texttt{maxWorkers} to limit the number of workers or use the \texttt{runInBand} option to have the test suites run sequentially.
Jest determines, based on previous runs, if it will be faster to run test suites sequentially rather than in parallel without informing the developer\cite{testSchedular}. 
It also orders the test suites to run based on past runtime, failure history, and/or file size (if no prior information is available)~\cite{yost2023finding}. 
Since these factors can change at any time, Jest may arbitrarily determine to run test suites sequentially and in some non-deterministic order, potentially leading to \odtest{}s that fail when Jest happens to run them in a failing order.

On the other hand, Jest runs tests inside test suites in the order they are encountered in the file (i.e., the first test in the file runs first).
If there are \describe{} \block{}s to group tests, Jest runs them in the order they are defined within the file~\cite{jestDocs}. 
Jest runs all the \describe{} \block{}s in a test suite before executing any actual tests~\cite{jestDocs}.

Jest has different options for test parallelization and selection~\cite{jestDocsCli}.
For example, the \texttt{testNamePattern} option allows running only individual tests or \describe{} \block{}s that match a specific naming pattern. 
The \texttt{shard} option allows splitting a test suite into multiple shards and then running specific shards, which is useful for parallelizing test execution across multiple machines or environments to reduce the overall test runtime (each shard is run on different machines).
The \texttt{randomize} option was released in late 2023 in response to a popular feature request from developers~\cite{jest_issue_4386}, who indicated the need for test randomization that is common in other testing frameworks such as JUnit and NUnit.
From the discussion of this feature, we observed that developers saw that randomizing test execution helps uncover cases where tests inadvertently depend on the state left by previous ones. 

\begin{figure*}[h]
    \centering
    \includegraphics[width=0.85\linewidth]{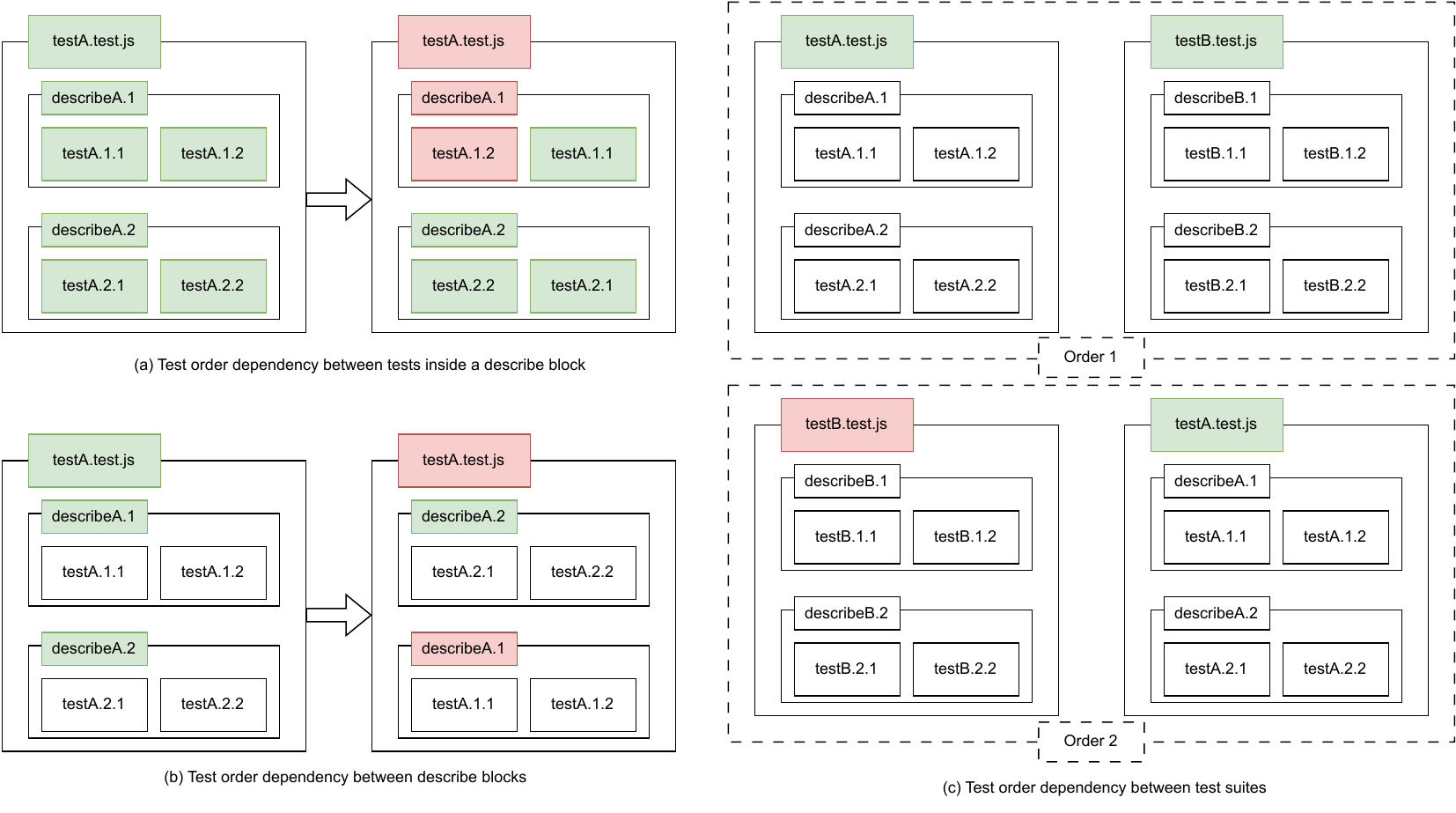}
    \caption{Different levels of test order dependency with passed tests in green and failed tests in red}
    \label{fig:level}
\end{figure*}

In Jest, test dependencies can manifest at various levels.
Order dependency may arise between tests, \describe{} \block{}s, and test suites.
Fig.~\ref{fig:level}(a) illustrates an example of order dependency between two tests.
Changing the order of \texttt{testA.1.1} and \texttt{testA.1.2} results in the failure of \texttt{testA.1.2}, as it depends on \texttt{testA.1.1}.
The order dependency between \describe{} \block{}s is demonstrated in Fig.~\ref{fig:level}(b), where running \texttt{describeA.2} before \texttt{describeA.1} leads to the failure of \texttt{describeA.1}.
Fig.~\ref{fig:level}(c) depicts order dependency between two test suites: \texttt{testA.test.js} and \texttt{testB.test.js}.
The passing order for these two tests is \texttt{testA.test.js} $\,\to\,$
 \texttt{testB.test.js}.
By changing the running order to \texttt{testB.test.js}$\,\to\,$
 \texttt{testA.test.js},  \texttt{testB.test.js} will fail.

\section{Approach}
\label{sec:method}


We implement JavaScript Test Order-dependency Detector (\jsTod{}) to detect \odtest{}s within JavaScript projects using Jest.
\jsTod{} first extracts all test paths from a given project and then reorders the tests and test suites using a user-defined number of \reorder{}s.
\jsTod{} runs all newly reordered tests for the given number of re-runs and finally exports the results (i.e., test outcomes) into JSON files.
\Space{We explain below the process that \jsTod{} follows to reorder and rerun tests.}

To reorder tests at all three levels (Fig.~\ref{fig:level}), 
we implement a randomization test reordering approach, shown in Algorithm~\ref{alg:reorder}.
The function \texttt{randomizeOrder} takes as input an array, representing the set of test suite paths, tests, or \describe{} \block{}s to be reordered, and an integer, representing the number of \reorder{}s.
The function returns the set of \reorder{}s.
If all possible permutations of items are smaller than the provided number of \reorder{}s, the function returns all permutations.
Otherwise, it shuffles the items to create a new \reorder{} and adds it to the set of \reorder{}s to be returned if it is unique.
This process repeats until reaching the input number of \reorder{}s. 

\begin{algorithm}
\caption{Tests reordering approach}\label{alg:reorder}
\begin{algorithmic}[1]
 \Function{randomizeOrder}{{array}, {reorder\_num}}
        \State $all \gets calcNumPermutations(array\_length)$
        \If{$reorder\_num\geq all$}
             \State $result \gets getPermutations(array)$
        \Else
            \While{$result\_length\leq reorder\_num$}
                \State $newArray \gets shuffleArray(array)$
                \If{$isUnique(result, newArray)$}
                    \State $result \gets newArray$
                \EndIf
            \EndWhile
        \EndIf 
    \State \Return {$result$}
\EndFunction

\end{algorithmic}
\end{algorithm}

\jsTod{} uses different approaches to ordering test suites, tests and \describe{} \block{}s\Space{, as detailed below}:

 \subsection{Reordering Test Suites}
For reordering test suites (Fig.~\ref{fig:level}(c)), we need to extract the paths of test suites from each project and pass the running order to Jest.
We extract the test suites' paths for a given project automatically using Jest's \texttt{-listTests} option~\cite{jestDocsCli}.

Then, we pass the test suite paths to the \texttt{randomizeOrder} function in Algorithm~\ref{alg:reorder} to generate up to \texttt{reorder\_num} unique \reorder{}s.
    In addition, Jest uses a \texttt{TestSequencer}\cite{testSequencer} 
    class to determine the order in which to run the test suites.
    We write a custom sequencer that extends this \texttt{TestSequencer} class, allowing us to run the test suites in a specific order. 
    Listing~\ref{lst:custom} shows our custom sequencer, which receives the running order as an argument and returns tests in the defined order.
    We provide the custom sequencer and running order to Jest by passing the \texttt{CustomSequencer} path after the \texttt{-testSequencer} option and order of test suites paths after the \texttt{-order} option, e.g., \texttt{npx jest --runInBand --testSequencer='/path/to/customSequencer.js'  --order='/path/to/test1.js,/path/to/test2.js'}.

    \begin{lstlisting}[caption={CustomSequencer class},label={lst:custom}]
class CustomSequencer extends TestSequencer {
 sort(tests) { //get the running order and put it in orderPath array
  const orderPathString = process.argv.find ((arg)=>arg.startsWith("--order")).replace("--order=","");
  const orderPath =orderPathString.split(","); 
  
  return tests.sort((testA, testB) => {
   const indexA =orderPath.indexOf(testA.path);
   const indexB =orderPath.indexOf(testB.path);
   if (indexA === -1 && indexB === -1) return 0; // Keep both tests in their original order if not specified in orderPath
   if (indexA === -1) return 1;
   if (indexB === -1) return -1;
   const result = indexA - indexB;
   return result
   ;})
 ;}}
    \end{lstlisting}

 \subsection{Reordering Tests and \describe{} \block{}s} Algorithm~\ref{alg:reorderD} describes the process for extracting and reordering both tests and \describe{} \block{}s (Fig.~\ref{fig:level}(a),(b)).
    After extracting test suites, we use the Babel toolchain~\cite{Babel·Ba89:online}, 
    a JavaScript compiler and source code transformer, to parse each test suite. Babel generates an Abstract Syntax Tree (AST) for each test suite (including nodes, parameters and types). 
    \Space{Multiple node types exist in Babel.
    }Using the AST, we identify all \describe{} \block{}s within each test suite of a given project.
    The \describe{} \block{}s are recognised by recording AST nodes when a node's type is ``\textit{identifier}" and its name is ``\textit{describe}".
    These blocks are then reordered and saved in new test files in the original directory. 
     For identifying tests inside each \describe{} \block{}, we use the same randomizing approach as for Algorithm~\ref{alg:reorderD},
     except the identifier names of these AST nodes are either \texttt{it} or \texttt{test}.

\begin{algorithm}
\caption{Extracting and reordering \describe{} \block{}s/tests in \jsTod{}}\label{alg:reorderD}
\begin{algorithmic}[1]
\\\Comment{Input: a project path}
\\\Comment{Output: results of running in JSON format saved in the project path.}
\State $INSTALL\_PROJECT(project\_path)$
\State $testSuites \gets ListTestSuites(project\_path)$
 \For{\texttt{testSuite in testSuites}}
   \State $blocks \gets PARSE(testSuite)$
   \If{$len(blocks)\geq 2$}
    \State $newTests \gets randomizeOrder(blocks,
      reorder\_num)$
    \For{\texttt{newTest in newTests}}
        \State $newTestCode \gets generateCode(newTest)$
        \State $name \gets generateName(newTestCode)$
        \State $save(name, newTestCode)$

    \EndFor
   \EndIf
 \EndFor
\For{\texttt{newTestSuite in newTestSuites}}
    \State $RUN(newTestSuite, rerun\_num)$
\EndFor

    \\

\Function{PARSE}{testSuite}
    \State $ast \gets babel\_parser(testSuite)$
    \State $blocks \gets TRAVERSE(ast, type)$ 
    \State \Return {$blocks$}
\EndFunction
\end{algorithmic}
\end{algorithm}

We record the results of each run in a JSON file for further investigation.
In cases where a test suite consistently fails across all reruns of the same order, we categorize it as an \odtest{}.
Otherwise, if it passes and fails even as the order remains the same, the test suite is flaky, depending on other non-deterministic factors.

\section{Experimental Setup}
\label{sec:experimentSetup}
\begin{figure}[!htp]
    \centering
    \includegraphics[width=\linewidth]{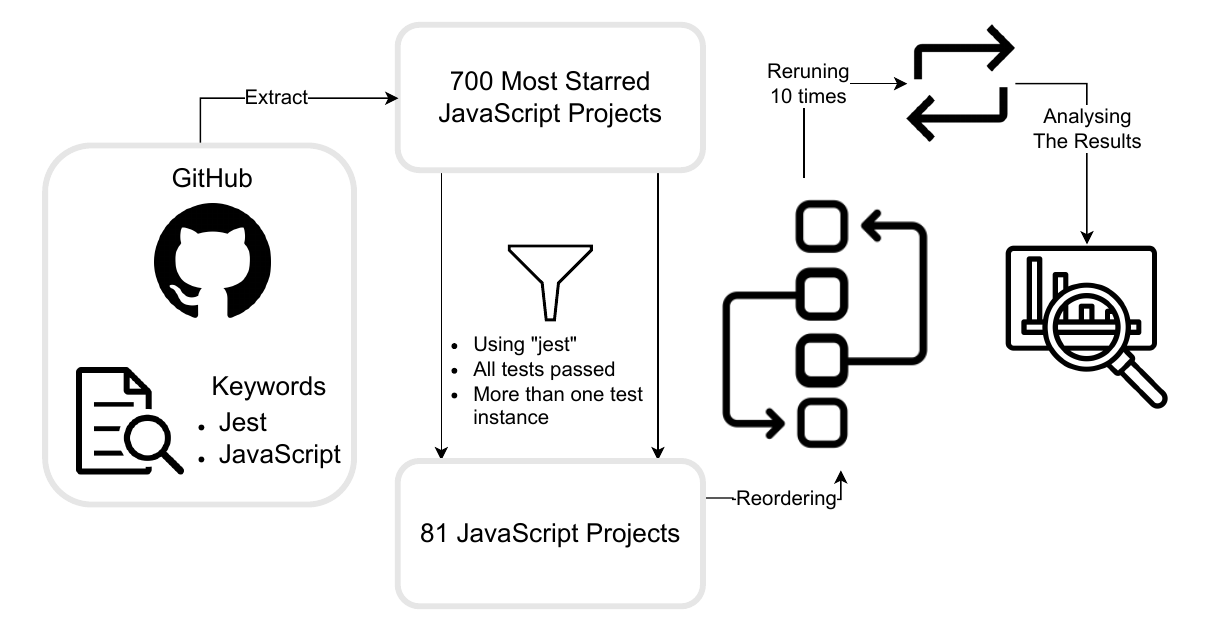}
    \caption{An overview of the data collection and analysis process}
    \label{fig:flowchart}
\end{figure}

Fig.~\ref{fig:flowchart} shows
an overview of our experiment, including data collection and analysis processes.
For our evaluation, we reorder each test suite, test, and \describe{} \block{} for \NumReorders{} \reorder{}s and reran each new test suite \NumReruns{} times using \jsTod{}.

\subsection{Dataset}
To collect a dataset of JavaScript projects for evaluation, we utilise GitHub's REST API~\cite{RESTAPIe8:online} 
to search for projects.
We use a keyword search to collect projects that contain "jest" in the repository (e.g., topics, issues, and README files).
In addition, we searched for JavaScript projects without any special keywords.
In both search queries, we order the projects based on GitHub's stars,\Space{ Users utilize GitHub's star feature to mark repositories they find valuable, reflecting a project’s level of interest to the community.}
where there is a correlation between the number of stars and the number of forks and contributors, demonstrating the project’s popularity~\cite{brisson2020we}.
Based on our search results, we select the \NumInitialProjects{} most starred projects to form a large enough initial dataset for inspection.
After cloning all \NumInitialProjects{} projects, we select the projects that use Jest as their testing framework by searching with the keyword \texttt{jest} in the \texttt{package.json} file.
We find \NumJestProjects{} projects that use Jest.

For each project, we execute all tests in the default order (i.e., the original test order as they appear in the project).
After each execution, we record the number of test suites, \describe{} \block{}s, and tests for further investigation.
We then exclude the projects that do not have enough test instances to investigate (a project should contain at least two tests so that we can reorder them). 
We find that \NumProjectsWithWorkingTests{} of the \NumJestProjects{} selected projects have more than one test instance (at least two tests, \describe{} \block{}s, or test suites) and all passing tests when run in the default order.
We exclude all other projects from our evaluation. 

When we run \jsTod{}, we find a project
(``snapshot-diff'') where the reordering does not work.
The project has a test with a missing semicolon that causes Babel to fail in transforming the tests.
Hence, we exclude this project from our evaluation and continue with the remaining \NumEvaluatedProjects{} projects.
Table~\ref{tab:dataset} shows summary statistics of the selected projects.

\begin{table}[!htp]\centering
\caption{Project Statistics}\label{tab:dataset}

\begin{tabular}{lrr|rrrr}\toprule
      & \multicolumn{2}{|c|}{Production}   & \multicolumn{4}{c}{Tests}  \\\cmidrule{2-7}
      & \multicolumn{1}{|r}{LOC}     & \multicolumn{1}{r|}{\#files}      & \multicolumn{1}{r}{LOC}     &\multicolumn{1}{r}{\#tests} &\multicolumn{1}{>{\raggedleft}p{1cm}}{\#describe blocks}& \multicolumn{1}{>{\raggedleft}p{1cm}}{\#test suites} 
      \\\midrule
Mean  & \multicolumn{1}{|r}{1791}    & 32           & 690     &79   &18   &11      \\
Total & \multicolumn{1}{|r}{145092}  & 2583         & 55911   &6398 &1420 &873     \\
\bottomrule
\end{tabular}

\end{table}

\subsection{Environment}
We conduct our evaluation on a machine with a 2.3 GHz Intel Core i9 processor with 16 GB RAM running on a macOS 14.
We use \texttt{npm} version 9.5.1 and \texttt{Node.js} version 18.16.1.
The experiment took approximately 121 hours to run across all projects and reruns required (see Table~\ref{tab:results} for runtime data). 

\subsection{Manual analysis}

We conduct two different manual analyses on the collected results.
We conduct our first manual analysis on the tests before reordering anything.
The goal is to check if all tests of a project pass by default, that all test paths are available and correctly extracted by Jest's \texttt{-listtests} option
(available from Jest version 20.0.0\cite{jestV20}). 
In eight projects, the  \textit{-listTests} option could not list the test paths due to using older versions of Jest.
In those projects, we manually extract the test paths to pass to \jsTod{}, allowing us to evaluate on those projects.

After reordering the tests, we conduct a second manual analysis to verify the cause of failures for failing tests.
One of the authors manually inspected all results, examining the source code and error messages to identify instances of failed orders.
After identifying the cause of failure, a second author independently inspected the code to confirm if it was due to an \odtest{}.
The authors discussed cases where there were disagreements, and in one case, a third author confirmed the cause of flakiness.
Our goal is to create a categorization of common reasons for flakiness, namely the types of shared state between tests, among the detected \odtest{}s.

We make publicly available our \jsTod{} implementation and a full replication package for our evaluation, which includes all data, scripts and full results~\cite{replicationPackage}.
\section{Results}
\label{sec:results}
We apply \jsTod{} on the tests within each of the \NumEvaluatedProjects{} \project{}s we evaluate, reordering tests at the three levels of tests, \describe{} \block{}s, and test suites. 
\Space{Already discussed previously concerning first manual analysis...
In applying \jsTod{},
we find that our implementation is accurate at determining the correct test paths, returning the correct test paths for 90\% of the projects (73 out of \NumEvaluatedProjects{} projects).
}


\subsection{RQ1. Manifesting \odtest{}s}
Table~\ref{tab:results} shows the results of applying \jsTod{} on each \project{}.
The total number of \project{}s in Table~\ref{tab:results} is the summation of all \project{}s in the three different levels that overlap with each other in some \project{}s.

In total, we find \NumODTests{} \odtest{}s across \NumProjectsWithODTests{} \project{}s.
All \NumODTests{} tests fail across all \NumReruns{} reruns of the same order, confirming that they indeed are \odtest{}s.
The majority of these \odtest{}s were found by reordering at the level of tests (\NumODTestsAtTestLevel{} tests).
Only \NumODTestsAtDescribeLevel{} \odtest{}s were found at the level of \describe{} \block{}s.

\begin{table*}[!ht]\centering

\caption{Results of Reordering and rerunning Tests in Three Levels}\label{tab:results}

\begin{threeparttable}

\begin{tabular}{p{1.8cm}r|rr|rrrrrrr}\toprule
                 & &\multicolumn{2}{c|}{Default Order} & \multicolumn{7}{c}{Reordering Result}\\\cmidrule{3-11}
Reordering Level & \multicolumn{1}{|p{1.2cm}|}{\#\project{}s}   & \multicolumn{1}{>{\raggedleft}p{1cm}}{\#tests}  & \multicolumn{1}{>{\raggedleft}p{1.2cm}|}{\#test suites} & \multicolumn{1}{>{\raggedleft}p{1cm}}{\#test} & \multicolumn{1}{>{\raggedleft}p{1.2cm}}{\#test suites} & \multicolumn{1}{>{\raggedleft}p{1.2cm}}{\#failed tests} & \multicolumn{1}{>{\raggedleft}p{1.3cm}}{\#failed test suites} & \multicolumn{1}{>{\raggedleft}p{1cm}}{\#ODFT\tnote{*}}  & \multicolumn{1}{>{\raggedleft}p{1cm}}{\#ODFTS\tnote{**}} & \multicolumn{1}{>{\raggedleft}p{1.3cm}}{runtime (s)\tnote{***}}    \\\midrule
\multicolumn{1}{l|}{Test}             & 67           & 6169     & 782           & 44713  & 4353          & 108            & 65         & \textbf{\NumODTestsAtTestLevel{}}         & \textbf{19}         & 273788.1 \\
\multicolumn{1}{l|}{Describe block}   & 32           & 5818     & 692           & 18534  & 558           & 11             & 6          & \textbf{3\Space{\NumODTestsAtDescribeLevel{}}}          & \textbf{1}          & 116013.6  \\
\multicolumn{1}{l|}{Test suite}       & 49           & 5960     & 850           & 59600  & 8500          & 0              & 0          & 0                   & 0                   & 44453.3  \\\midrule
\multicolumn{1}{l|}{Total}            & 148          & 17947    & 2324          & 122847 & 13412         & 119            & 71         & \textbf{\NumODTests{}}         & \textbf{20}         & 434255.0 \\

\bottomrule
\end{tabular}
\begin{tablenotes}
      \small
      \item[*] order-dependent flaky tests
      \item[**] order-dependent flaky test suites 
      \item[***] runtime calculated for the total of 10 reorders each executed 10 times (100 runs in total)
       
    \end{tablenotes}

\end{threeparttable}

\end{table*}

\subsubsection{Order dependency between tests}
 
From all \NumEvaluatedProjects{} \project{}s, we find that \NumProjectsWithSingleTestInDescribe{} of those \project{}s contain only a single test inside each \describe{} \block{}, and thus, we do not include them in our analysis of tests.
Among others, running \jsTod{} on \project{} ``jest-serializer-vue'' was too time-consuming (more than six hours)\Space{ to reorder and rerun tests}, and ``draggable'' has a test with a missing semicolon that causes Babel to fail in transforming the test code. 
Therefore, we exclude these \NumProjectsToExclude{} \project{}s and continued with the remaining  \NumFinalProjectsForAnalysis{} \project{}s for the test-level analysis.

\Fix{JS-TOD successfully reorders the tests for 85\% of all projects (57 out of 67 projects), with the remaining projects fixed manually.} Our analysis reveals a total of \NumODTestsAtTestLevel{} \odtest{}s from nine \project{}s.
As shown in Table~\ref{tab:tests}, most of the flaky tests appear in \project{} ``jest-webextension-mock'' with 35 \odtest{}s, followed by ``serverless-jest-plugin'', which has five \odtest{}s, and ``native-testing-library'' with four \odtest{}s. The rest of the six \project{}s contain one or two \odtest{}s. 

\subsubsection{Order dependency between \describe{} \block{}s}
\Space{Here, we investigate whether there are any order dependencies between different \describe{} \block{}s that can result in flakiness.}
We notice that 47 out of the \NumEvaluatedProjects{} \project{}s have no or just a single \describe{} \block{} inside each test suite, and therefore, it is not possible to reorder the \describe{} \block{}s for such \project{}s.
\Space{For example, ``electron-typescript-react'' has only a single \describe{} \block{} in each test suite.
}In addition, ``draggable'' has a test with a missing semicolon that causes Babel to fail in transforming the \project{}'s tests.
As a result, we continue with the 33 \project{}s for further analysis.

In some cases, the \describe{} \block{}s can be nested (i.e., a \describe{} \block{} can contain other \describe{} \block{}s within its structure). For the 33 remaining \project{}s, we reorder the \describe{} \block{}s up to two nested levels.
One \project{}, ``Jasmine-Matchers''~\cite{Jasmine-Matchers}
has more than two nested blocks, and we could not find a suitable way to reorder its blocks given technical challenges in rewriting those tests, so we exclude this \project{} from our analysis.
Ultimately, we analyse 32 \project{}s.

Among the 32 \project{}s, we find one \project{}, ``jest-build-ins'', with two order-dependent \describe{} \block{}s (Table~\ref{tab:describe}), namely ``native jest mock function inspectors'' and ``exposing mock internals''.

\subsubsection{Order dependency between test suites}\label{sec:RQ1}

We only analyse \project{}s with multiple test suites to determine order dependency between test suites, resulting in
62 out of \NumEvaluatedProjects{} \project{}s. 
\Space{As explained in Section~\ref{sec:method}, we introduced the sequencer path to Jest using our custom sequencer class and the \texttt{-testSequencer} option, which is available from Jest version 24.7.0\footnote{\url{https://github.com/jestjs/jest/releases/tag/v24.7.0}}.
}In 13 of the 62 \project{}s, we encountered an issue where the \texttt{-testSequencer} was unrecognized in the \texttt{npx} command due to using older versions of Jest.
Since we rely on  Jest's test sequencer class to enforce a specific running order, we cannot reorder the tests for these \project{}s.
All test suites pass for the remaining 49 \project{}s regardless of their test-suite order.
Thus, we do not find any order dependency between test suites. 



\subsection{RQ2. Causes of \odtest{}s}
We present the results of our root-cause analysis on the detected \odtest{}s at the test level and at the \describe{} \block{} level.

\subsubsection{Order dependency between tests}\label{sec:reorderTestsResults}
 

\begin{table*}[!ht]\centering
\caption{Results of Reordering Tests in 67 \Project{}s }\label{tab:tests}
  \begin{threeparttable}

\begin{tabular}{p{3.3cm}|rr|rrrrrrlr}\toprule
                                 & \multicolumn{2}{c|}{Default Order} & \multicolumn{8}{c}{Reordered tests} \\\cmidrule{2-11}                                                                          
\Project{}                     & \multicolumn{1}{l}{\#tests}& \multicolumn{1}{>{\raggedleft}p{0.7cm}|}{\#tests suites} & \multicolumn{1}{l}{\#test}  & \multicolumn{1}{>{\raggedleft}p{0.7cm}}{\#test suites} & \multicolumn{1}{>{\raggedleft}p{0.7cm}}{\%failed test}  & \multicolumn{1}{>{\raggedleft}p{1.3cm}}{\%failed tests suites} & \multicolumn{1}{l}{\#ODFT} & \multicolumn{1}{l}{\#ODFTS} & \multicolumn{1}{p{1.5cm}}{OD cause} & \multicolumn{1}{>{\raggedleft}p{1.2cm}}{runtime(s)} \\\midrule
jest-image-snapshot         & 100   & 5   & 980  & 46  & 5.0\%            & 35\%                & \textbf{2}  & \textbf{1} & file sharing & 12404.5  \\
jest-junit                 & 53    & 6   & 496  & 34  & 7.0\%            & 18\%                & \textbf{1}  & \textbf{1} & file sharing & 1201.6   \\
\rowcolor{light-gray}jest-when\tnote{\#}                & 86    & 1   & 860  & 10  & 1.0\%            & 100\%               & 0           & 0          & - & 820.1    \\
jest-webextension-mock      & 173   & 14  & 391  & 96  & 38.0\%           & 68\%                & \textbf{35} & \textbf{7} & shared mocking state & 7491.8   \\
\rowcolor{light-gray}commander.js\tnote{\#}             & 1108  & 102 & 10189& 867 & 0.1\%          & 1\%                 & 0           & 0          & - & 54987.3  \\
jest-mock-now               & 5     & 1   & 46   & 6   & 8.7\%          & 67\%                & \textbf{1}  & \textbf{1} & file sharing & 411.8    \\
KaTeX                      & 1216  & 8   & 8020 & 62  & 0.2\%          & 16\%                & \textbf{2} & \textbf{1}  & file sharing & 8174.6   \\
native-testing-library      & 159   & 36  & 949  & 196 & 2.0\%            & 4\%                 & \textbf{4} & \textbf{1}  & shared mocking state& 14322.9  \\
react-testing-library      & 237   & 14  & 392  & 75  & 0.3\%          & 1\%                 & \textbf{1}  & \textbf{1} & file sharing & 2329.3   \\
serverless-jest-plugin     & 30    & 6   & 288  & 52  & 10.0\%           & 54\%                & \textbf{5} & \textbf{5}  & file sharing & 6507.7   \\
vue-testing-with-jest-conf17 & 11   & 3   & 78   & 14  & 6.0\%            & 36\%                & \textbf{1}  & \textbf{1} & file sharing& 544.6    \\\midrule
Total	&3178	&196	&22689	&1458	&&		&\NumODTestsAtTestLevel{}	&19	& &109196.2\\
\bottomrule
\end{tabular}
 \begin{tablenotes}
      \small
      \item[\#] shaded rows are \project{}s with failing tests but not order-dependent flaky tests
       
    \end{tablenotes}
\end{threeparttable}

\end{table*}

Table~\ref{tab:tests} shows the results of our analysis on detected \odtest{}s at the test level.
In total, 56 \project{}s have all tests passing across all \NumReorders{} \reorder{}s for \NumReruns{} reruns, while 11 \project{}s experienced at least one failed test after being reordered (in one of the \NumReruns{} reruns).
We manually investigate the results of those 11 \project{}s to identify the cause of test failures. 
We identify \NumODTestsAtTestLevel{} \odtest{}s from the nine \project{}s that have them (shaded rows in Table~\ref{tab:tests}).

We find the causes for the \odtest{}s can be divided into two categories: sharing file and mocking state.
Among the detected \odtest{}s, we identify \NumODTestsFiles{} tests whose failures are caused by file sharing across seven \project{}s, and \NumODTestsMocks{} tests caused by shared mocking state between tests across two \project{}s.

\noindent \textbf{File sharing:} We find that developers use different methods to share a file between tests.
For example, 
we identify one \odtest{} within the ``buildJsonResults'' test suite in ``jest-junit'', which is related to two tests: ``should report no results as error'' and ``should honour templates when the test has errors''.
The two tests shared a mock file (as seen in Listing~\ref{lst:junit} - lines 2 and 7).
The first test sets \texttt{reportTestSuiteErrors: "true"} on line 4, which alters the cached object of the mock file used by \texttt{require} (lines 2 and 7) and consequently impacts the outcome of the subsequent test.
On line 11, the test checks whether the total number of errors is equal to one.
In case the second test runs first, the \texttt{totals.errors} becomes not equal to one, 
therefore the test fails.
Consequently, we notice that whenever ``should honour templates when the test has errors'' executes before ``should report no results as error'', the latter test fails.
Hence, ``should honour templates when the test has errors'' is a \polluter{} and ``should report no results as error'' is a \victim{}. 

\begin{lstlisting}[caption={Tests that shared a file from jest-junit},label={lst:junit}]
it('should honor templates when test has errors', () => {
    const failingTestsReport = require('../__mocks__/failing-compilation.json');
    ...
    reportTestSuiteErrors: "true",
    ...
it('should report no results as error',() => {
    const failingTestsReport = require('../__mocks__/failing-compilation.json');
    ...
    reportTestSuiteErrors: "true"
    ...   
    expect(totals.errors).toEqual(1);
  
\end{lstlisting}

Another example of file sharing\Space{ that leads to order-dependent flakiness} is in ``jest-image-snapshot''.
Some of the tests create/update image snapshot files accessed by other tests.
\Space{The location of these files (obtained using \texttt{getSnapshotFiles()}) is shared by the tests (shown in Listing~\ref{lst:image}).
}Specifically, the test ``should write the image snapshot on first run'' writes files to the shared location.
The tests ``should not delete the snapshot when environment flag is not enabled'' and ``should delete the snapshot when environment flag is enabled'' expect those files to exist.
That is, the latter two tests depend on the state set up by the first. 

\begin{lstlisting}[caption={Tests with shared files between them from jest-image-snapshot},label={lst:image}]
function getSnapshotFiles() {
 return fs.readdirSync(path.join(tmpDir, '__image_snapshots__'));}
it('should write the image snapshot on first run',()=>{
 const{status, stdout, stderr}=runJest(['-u']);
 expect(stderr).toContain('snapshots written');
 expect(status).toEqual(0);
 expect(stdout).toEqual('');
 expect(getSnapshotFiles()).toHaveLength(3);});
it('should not delete the snapshot when environment flag is not enabled', () => {
 const { status, stdout } = runJest(['-u', 'image.test.js']);
 expect(status).toEqual(0);
 expect(stdout).toEqual('');
 expect(getSnapshotFiles()).toHaveLength(3);});
it('should delete the snapshot when environment flag is enabled', () => {
 const { status, stdout, stderr } = runJest(['-u', 'image.test.js'], {
  JEST_IMAGE_SNAPSHOT_TRACK_OBSOLETE: '1',});
 expect(stderr).toContain('outdated snapshot');
 expect(status).toEqual(0);
 expect(stdout).toEqual('');
 expect(getSnapshotFiles()).toHaveLength(1);});
\end{lstlisting}

\noindent \textbf{Shared mocking state:}
Mocking states is a type of state sharing that specifically looks at the state of a mock.
One example of sharing mocking states is ``jest-webextension-mock''.
The \project{} has seven test suites with 35 \odtest{}s. All 35 tests fail after reordering those tests due to the issue of checking the number of times functions are called.

An example from the \project{} is shown in Listing~\ref{lst:mock}. The ``command'' test suite includes two tests: ``getAll'' and ``getAll promise''.
In ``getAll'', when the function \texttt{browser.commands.getAll()} is called (line 3), the \texttt{expect} statement anticipates the function to be called once (line 4).
The ``getAll promise'' test calls this function (line 7).
If ``getAll promise'' runs before ``getAll'', the function gets called twice, causing the \texttt{expect} statement in line 4 to return false and a test to fail. A possible fix is to use \texttt{jest.clearAllMocks()} to clear the mocking state in a \texttt{beforeEach} hook, which ensures that the mocks are cleared and the mock state is not shared between tests.

\begin{lstlisting}[caption={Tests with shared mocking state from jest-webextension-mock},label={lst:mock}]
  test('getAll', (done) => {
   ...
    browser.commands.getAll(callback);
    expect(browser.commands.getAll).toHaveBeenCalledTimes(1);
    ...
  test('getAll promise', () => {
    return expect(browser.commands.getAll()).resolves.toBeUndefined();
\end{lstlisting}


Listing~\ref{lst:native} shows another example of \odtest{}s caused by a shared mocking state. The test suite has five tests, of which four are expected to use the mock function on line 2. The remaining test,  ``async act should not show an error when ReactTestUtils.act returns something'', uses the same mock function with a new and different implementation specific to the test (line 12). In the default order, test ``async act should not show an error when ReactTestUtils.act returns something'' is the last test in the test suite, meaning it will run last, and therefore it does not share its inner mocking function with other tests. As a result of reordering tests, the \texttt{jest.mock(`react-test-renderer')} defined in this test will impact the other four tests as the mocking state changes, which causes the other tests that come after the target test to fail.

\begin{lstlisting}[caption={Tests with shared mocking state from native-testing-library},label={lst:native}]
    let asyncAct;
jest.mock('react-test-renderer', () => ({
  act: cb => { ...}}))
beforeEach(() => {
  jest.resetModules();
  asyncAct = require('../act-compat').asyncAct;
  jest.spyOn(console,'error').mockImplementation(()=>{});})
afterEach(() => {
  console.error.mockRestore();})
test('async act should not show an error when ReactTestUtils.act returns something', async () => {
  jest.resetModules();
  jest.mock('react-test-renderer',()=>({...
\end{lstlisting}



\noindent \textbf{Other failed tests:} We observe two \project{}s with failing tests after the reordering process: ``jest-when''~\cite{timkindb36:online} and ``commander.js''~\cite{tjcomman9:online}, but we do not find the failing tests to be \odtest{}s.
In "jest-when," we observe test logic that depends on the test name and file structure.
For example, as shown in Listing~\ref{lst:when}, test ``fails verification check if all mocks were not called with line numbers'' (line 5) depends on the name of the test suite.
The tests fail when we transform them into new suites (i.e., the newly reordered test files).
Therefore, we attribute the failure to the structure of the test code and not test order dependency.
In addition, ``commander.js'' experiences failed tests when rerun in some orders.
After analysis, we discovered these failures are false positives, more related to resource issues than test order dependency.

\begin{lstlisting}[escapeinside={(*@}{@*)},caption={Tests from jest-when},label={lst:when}]
    try {
    verifyAllWhenMocksCalled();
  } catch (e) {
    const errorLines = e.message.split('\n');
    const currentFilePathPattern = /src(?:\\|\/)when\.test\.js:\d{3}(.|\s)*/;
    
\end{lstlisting}

There were also nine \project{}s 
with failing test suites after reordering the tests.
After manually analyzing the results, we find that the failures occur due to Babel not parsing some parts of the original test suites, and thus, \jsTod{} ends up not adding them to the new test suites.
For example, Listing~\ref{lst:jest} from ``jest-it-up''~\cite{rbardini2:online} 
shows part of a test suite that has a \texttt{require} statement in line 2.
\jsTod{} does not identify this statement when traversing the AST of the original test suite and does not include it in new test suites after reordering.
This parsing error causes the test "returns the new contents, and changes if new thresholds" to fail.
As a result, we manually fix the code of the new test suites by adding the missing code segments and then comparing those with the original suites to ensure that it is identical.
After this manual modification, we rerun all test suites, resulting in all tests passing. 

In addition, there are two \project{}s (``jest-fail-on-console'' and ``KaTeX'') with specific Jest configurations for running tests with defined name patterns.
As we add reordered tests to the new test suite, we change Jest's configuration to run them.

\begin{lstlisting}[escapeinside={(*@}{@*)},caption={Example of unidentifiable code by AST from jest-it-up},label={lst:jest}]
const fs = require('fs')
require('ansi-colors').enabled = false
    ...
it('returns the new contents and changes if new thresholds', () => {...  
  \end{lstlisting}

\subsubsection{Order dependency between \describe{} \block{}s}

Table~\ref{tab:describe} shows the \project{}s with failing tests after reordering \describe{} \block{}s. 
The results show that all reordered \describe{} \block{}s passed for all \project{}s except three: ``jest-when'', ``commander.js'' and ``fetch-mock-jest''.
As mentioned in Section~\ref{sec:reorderTestsResults}, ``jest-when'' has a test that relies on the names and structures of the test suites.
Failing tests in ``commander.js'' are caused by resource-related issues.
Hence, we do not attribute the failed tests to order-dependent flakiness.

``fetch-mock-jest''~\cite{fetch-mock-jest} 
has two order-dependent \describe{} \block{}s.
The \describe{} \block{}s contain three tests that expect a function to be called a set number of times (lines 5, 9 and 13 in Listing~\ref{lst:fetch}).
As a result of this shared mocking state, changing the orders of \describe{} \block{}s leads to tests failing.

\begin{lstlisting}[escapeinside={(*@}{@*)},caption={\describe{} \block{}s from fetch-mock-jest},label={lst:fetch}]
describe('exposing mock internals', () =>{
    ...
 it('exposes `calls` property', () =>{
	expect(fetch.mock.calls).toBeDefined();
	expect(fetch.mock.calls.length).toBe(2);
	...
 it('exposes `results` property', async () =>{
	expect(fetch.mock.results).toBeDefined();
	expect(fetch.mock.results.length).toEqual(2);
	...
describe('native jest mock function inspectors', () => {
 it('.toHaveBeenCalled()', () => {
    expect(() => expect(fetch).toHaveBeenCalled()).not.toThrow();
\end{lstlisting}

\begin{table*}[!htp]\centering
\caption{Result of Reordering \describe{} \Block{}s}\label{tab:describe}
\begin{threeparttable}
    
\begin{tabular}{p{2.5cm}|rr|rrrrrrlr}\toprule
                                 & \multicolumn{2}{c|}{Default Order} & \multicolumn{8}{c}{Reordered tests} \\\cmidrule{2-11}                                                                          
\Project{}                     & \multicolumn{1}{l}{\#tests}& \multicolumn{1}{>{\raggedleft}p{0.7cm}|}{\#tests suites} & \multicolumn{1}{l}{\#test}  & \multicolumn{1}{>{\raggedleft}p{0.7cm}}{\#test suites} & \multicolumn{1}{>{\raggedleft}p{0.7cm}}{\%failed test}  & \multicolumn{1}{>{\raggedleft}p{1.3cm}}{\%failed tests suites} & \multicolumn{1}{l}{\#ODFT} & \multicolumn{1}{l}{\#ODFTS} & \multicolumn{1}{p{1.5cm}}{OD cause} & \multicolumn{1}{>{\raggedleft}p{1.2cm}}{runtime(s)} \\\midrule

jest-when       & 86           & 1                  & 172    & 2             & 1.0\%            & 100.0\%                & 0                & 0          & -       & 417.4                   \\
commander.js    & 1108         & 102                & 2590   & 122           & 0.2\%          & 2.9\%                & 0                & 0          & -       & 14616.0                 \\
\rowcolor{light-gray}fetch-mock-jest\tnote{\#} & 216    & 9  & 2076   & 76   & 0.4\%          & 5.3\%                & \textbf{3}       & \textbf{1} & shared mocking state       & 4320.0 \\\midrule
Total	&1410	&112	&4838	&200	&&		&3	&1	& &19353.4\\

\bottomrule
\end{tabular}
\begin{tablenotes}
      \small
      \item[\#] shaded rows are \project{}s with order-dependent flaky tests
     
    \end{tablenotes}
\end{threeparttable}
\end{table*}

\section{Discussion}
\label{sec:discusion}

\subsection{Order dependency flakiness in JavaScript:} 
Our investigation of order-dependent flakiness in JavaScript reveals only \NumODTests{} \odtest{}s from \NumProjectsWithODTests{} \project{}s. 
We observe a notable low number of \odtest{}s in JavaScript \project{}s compared to other studied languages, particularly Python~\cite{gruber2021empirical,wang2022ipflakies} and Java~\cite{luo2014empirical,lam2019idflakies}, where \odtest{}s are among the most prevalent types of flaky tests\Space{ in \project{}s written in those languages}.
Our results confirm findings from an empirical study of flakiness in JavaScript projects from issue tracking data, which also found very few cases of \odtest{}s in JavaScript projects~\cite{Hashemi2022flakyJS} (across multiple testing frameworks, including Jest).
We attribute this low number of \odtest{}s to two main factors: 1)~the way that JavaScript developers organize their tests when using Jest (that is, relevant tests are grouped together within the same blocks within the same test file), and 2)~the way that testing frameworks like Jest handle test order (running test suites in parallel by default or the test and \describe{} \block{}s in the order they appear in the file). 

There are no major factors in JavaScript that distinguish it from other languages in terms of how it can cause test order dependency.
In Java, \odtest{} flakiness caused by state-sharing is mostly due to modifying shared static fields~\cite{bell2014unit}.
Since the release of version ES6, JavaScript has provided support for static fields in classes and modules that can export properties.
Prior to ES6, developers used the prototype property to emulate them.
Our study does not reveal any \odtest{}s due to state sharing through these mechanisms.
This result could be due to how programmers write JavaScript code (i.e., in a more functional style) or the use of these features in programs.
Future work can investigate the types of state sharing in relation to language features.

Test runners ensure the ordered execution of tests in test classes/suites.
For Java projects, the JUnit testing framework can execute tests concurrently in the same JVM process.
Jest, on the other hand, runs tests in separate processes rather than in the same process, due to JavaScript being single-threaded by nature.
Jest uses node worker threads or child processes to parallelise test executions~\cite{JestRunningTests:online}.

\subsection{Implications} 
This study has implications for both the JavaScript developers/testers (who write test code) and designers of test management tools.
The results show that \odtest{}s manifest largely between tests rather than test suites or \describe{} \block{}s. 
While Jest by default ensures tests in the same test suite run in the declared order, developers are still keen to ensure test independence and even requested Jest to introduce a randomization feature to stress their testing~\cite{jest_issue_4386}.
Our implementation in \jsTod{} not only randomizes the order of tests in a test suite but also between \describe{} \block{}s and test suites, giving greater exploration options for developers to use.
Indeed, we even detected some dependencies between \describe{} \block{}s in our evaluation.
\Space{To ensure that there are no test-order dependencies between tests, developers may use our suggested randomization approach, implemented in \jsTod{}, to run tests in different orders.}    

Similarly to previous studies done on Java and Python~\cite{luo2014empirical,gyori2015reliable,gruber2021empirical,gruber2023flapy,zhang2014empirically}, we observe that \odtest{}s share states via external resources such as files.
Testers should be vigilant about reusing files between tests.
However, we find another major reason for \odtest{} flakiness is the sharing of mocking state, which was not discussed as a major reason in past studies for other languages.
In Jest, modules mocked using \texttt{jest.mock} function are intended to be mocked within the test suite that calls the mocking function~\cite{TheJestO19:online}, so their state persists between tests unless reset with \texttt{jest.resetModules()} or manually unmocked.
Although \texttt{jest.resetModules()} resets the module registry, it does not automatically reset all mocks that were applied globally.
\texttt{jest.resetModules()} only resets the module cache, meaning any new \texttt{require()} calls will reload the module.
However, if a module is mocked globally using \texttt{jest.mock()}, it still returns the mocked version, because the mock was applied outside the scope of the individual test. 


\Fix{We have conducted several experiments to explore ways to fix and mitigate flaky \odtest{}s, especially those caused by the shared mocking state. We found that one possible strategy for fixing shared mocking states is to ensure that each test starts with a fresh and clean mock state. To achieve this, one must explicitly reset the mock functions before or after each test. In the example provided in Listing~\ref{lst:native}, resetting \texttt{react-test-renderer} in the \texttt{beforeEach} block can help avoid the flakiness as it cleans the mock before each test. Additionally, calling the \texttt{jest.clearAllMocks()} function before each test can effectively clear mocks before execution. By clearing all mocking states in a \texttt{beforeEach} block, we successfully fixed 34 out of the 39 \odtest{}s that are caused by shared mocking states.}


Similar to most other testing frameworks, Jest also provides test fixture features (setup and teardown functions that get called before and after tests, respectively) that developers should use to eliminate \odtest{} issues, ensuring file and mock states are reset between tests.
\section{Threats to Validity}
\label{sec:threats}
\Space{There are several validity threats to the results of this study, which we discuss below.}

\noindent \textit{Number of runs and orders:} Deciding on how many reruns are required to reveal if a test is flaky has been pointed out as a significant issue in test flakiness research~\cite{amjed2022review}.  We decided to use 10 reorderings and 10 reruns for each new order (i.e., 100 reruns for each test suite). Based on our pilot runs and our previous experience from running flaky tests, we noted that with 10 reruns, we can observe the same results as increasing the number of reruns to 20 or 30. While increasing the number of reorders and reruns can potentially lead to finding more \odtest{}s, our experimentation with increased reorders and reruns showed no changes to the results. 

\noindent \textit{Reproducibility of results:} Our results from each program may not be easily reproducible as we used a randomization approach to reorder tests (i.e., each time we reorder the tests, we may get a different order). However, it is possible to reproduce the same results by running the same order we found to be flaky (see our replication package for more details~\cite{replicationPackage}). 

\noindent \textit{Generalizability of results:} The results are based on analyzing JavaScript programs that use Jest as their testing framework. We cannot guarantee that the results will hold for JavaScript programs that use other testing frameworks. Jest is the most popular framework and thus is considered representative. While there are similarities between the testing frameworks, one must adjust the approach using other testing frameworks, which requires updating the tool to the framework's options and settings. For example, not all testing frameworks have the concept of \describe{} \block{}s, so they would not have order-dependency between these \block{}s as we observed in Jest.

\noindent\Fix{\noindent \textit{Manual analysis:} Part of the project selection and the categorization of causes was done manually, which may affect construct validity due to oversight and bias. Two authors independently examined the source code and error messages to reduce the likelihood of false positives when categorizing the cause of flakiness. Whenever there was disagreement, we implemented a conflict resolution step where a third author was then involved to independently classify the disagreement. For all cases, all three authors rigorously discussed the causes until 100\% agreement was reached.}
\section{Related Work}
\label{sec:related}

Previous work has investigated various aspects of flaky tests, such as root causes and ways to identify, detect, predict, or mitigate flaky tests~\cite{thorve2018empirical,eck2019understanding,amjed2022review}. Several previous studies on test flakiness have largely focused on specific causes, such as test order dependency~\cite{gambi2018practical}, async wait and timing dependency~\cite{rahman2024flakesync,rahman2024flakerake}, concurrency~\cite{dong2020concurrency}, unordered collections~\cite{shi2016nondex}, random numbers~\cite{dutta2020detecting,dutta2021fixing}, code instrumentation~\cite{rasheed2023effect} or network flakiness~\cite{dietrich2022flaky}.
Of those cases, test order dependency has been emphasized as a major cause of test flakiness. The earlier work of Luo et al.~\cite{luo2014empirical} found that test order dependency is a major cause of flakiness in the analyzed Apache Java projects. 
Similarly, Gruber et al.~\cite{gruber2021empirical} reported that test order dependency is the main cause of flakiness in Python programs. 

\subsection{Order dependent test flakiness}
Several studies have investigated test order dependency in different contexts~\cite{lam2019idflakies,wang2022ipflakies,Shi2019iFixFlakies,gruber2023flapy,wei2021probabilistic,li2023systematically,li2022repairing,li2022evolution}.
Lam et al.~\cite{lam2019idflakies}  developed a framework, iDFlakies, which detects and classifies Java tests into order-dependent and non-order-dependent by running them in randomized orders. They found that 50.5\% of flaky tests in their study were caused by test order deficiency. IncIDFlakies~\cite{li2022evolution} improved upon iDFlakies by analysing code changes to detect newly introduced order-dependent flaky tests. The tool identifies and selects only tests with shared states between them, as such tests can potentially result in flakiness due to test-order dependency. 

Shi et al. proposed iFixFlakies~\cite{Shi2019iFixFlakies} to automatically fix \odtest{}s in Java. It relies on iDFlakies~\cite{lam2019idflakies} to detect \odtest{}s. Then, it reruns them in certain orders to determine whether an \odtest{} is a \victim{}, \polluter{} or \brittle{}. iFixFlakies searches for cleaners, which are tests that clean polluted state between tests, and uses their code to generate patches for repairing \odtest{}s.
Wang et al. used the same approach in iPFlakies~\cite{wang2022ipflakies} for detecting and fixing \odtest{}s in Python.
Li et al.~\cite{li2022repairing} later proposed ODRepair, an automated technique to repair \odtest{}s by determining the polluted shared state that results in the \odtest{} failing. It then generates code that can modify and reset the shared state so that the \odtest{} can pass.


Gruber et al.~\cite{gruber2023flapy} developed FlaPy, a tool to repeatedly execute the test suites of a given set of Python projects. FlaPy can categorize tests into four groups: not flaky, non-order-dependent, order-dependent and infrastructure.
Parry et al.~\cite{parry2022evaluating} used a machine-learning approach to classify tests into two groups: order-dependent and non-order-dependent. They experimented with the test suites of 26 Python projects to evaluate whether static and dynamic features were more effective as flakiness predictors than previous feature sets. 

\subsection{Flaky tests in JavaScript}
Hashemi et al.~\cite{Hashemi2022flakyJS} investigated flaky test causes and the actions taken by developers to fix or mitigate flaky tests in JavaScript projects. 
The study identified concurrency, async wait, OS and network as the top four causes of test flakiness in JavaScript projects. Unlike previous empirical studies on Java~\cite{luo2014empirical,eck2019understanding} and Python~\cite{gruber2021empirical},  which identify test order dependency as one of the key causes of flakiness, Hashemi et al.~\cite{Hashemi2022flakyJS} found only very few flaky tests are caused by test order dependency. 
Similarly, Costa et al.~\cite{costa2022test} investigated flaky tests across five programming languages, including JavaScript. 
The study reported that \emph{async wait}, \emph{concurrency}, and \emph{platform dependency} are the top flakiness causes in JavaScript.

Yost~\cite{yost2023finding} leveraged stress testing and reordering test suites to uncover flaky tests in JavaScript programs. By subjecting the application to stress conditions and modifying the order of test suite execution, the aim was to reveal the hidden issues that might not surface under normal circumstances. This study differs from ours in that they reordered test suites 10 times and ran each test suite only once, as well as also trying to introduce machine stress for non-order-dependency flakiness. 

FlakyFix~\cite{fatima2024flakyFix} used large language models to predict flaky test fix categories. To predict fixed categories, a machine-learning model analysed test code and identified patterns that lead to flakiness. They used two code models, CodeBERT and UniXCoder, and pre-trained them in six programming languages, including JavaScript.

Current flaky test detection tools are mainly available for Java and Python~\cite{amjed2022review}. Looking at previous reviews on tools that are used to detect flaky tests in JavaScript programs~\cite{parry2021survey,zolfaghari2021root,amjed2022review}, we identified only two tools. King et al.~\cite{king2018towards} proposed a machine-learning approach to detect flaky tests. They model the flakiness problem as a disease with certain symptoms to facilitate the identification of its causes. The tool uses Bayesian networks to classify and predict flaky tests.
NodeRacer~\cite{endo2020noderacer} can help detect test flakiness caused by event races in Node.js applications. NodeRacer comprises three main phases, collecting information, inferring the relation between the callbacks and rerunning the application.  It reruns the application several times to expose event race errors. NodeRacer specifically handles concurrency issues and does not address flaky tests directly.



\section*{Conclusion}
\label{sec:conclusion}

In this paper, we systematically investigate test flakiness due to test order dependency in JavaScript projects using Jest. We developed a randomization approach, called \jsTod, that targets test dependencies at three levels: tests, \describe{} \block{}s and test suites. \jsTod{}~ creates 10 randomised orders for each test item across the three categories, and for each new order, rerun each test suite 10 times.  
Our evaluation of \NumEvaluatedProjects{} projects shows that order-dependent test flakiness is possible in JavaScript projects.
Among 873 investigated test suites, we found \NumODTests{} \odtest{} instances in 20 test suites across 10 projects. All of the \odtest{}s we found are caused by shared files \Fix{(seven projects)} or shared mocking state \Fix{(three projects)} between tests. The latter is a new cause of \odtest{}s that has not been reported previously in the context of other languages.

\section*{Acknowledgment}
We acknowledge the support of the New Zealand SfTI National Science Challenge grant no. MAUX2004, the US National Science Foundation grant no. CCF-2145774 and CCF-2217696, and the Jarmon Innovation Fund.

\balance
\bibliographystyle{IEEEtran}
\bibliography{main}

\begin{thebibliography}{10}
\providecommand{\url}[1]{#1}
\csname url@samestyle\endcsname
\providecommand{\newblock}{\relax}
\providecommand{\bibinfo}[2]{#2}
\providecommand{\BIBentrySTDinterwordspacing}{\spaceskip=0pt\relax}
\providecommand{\BIBentryALTinterwordstretchfactor}{4}
\providecommand{\BIBentryALTinterwordspacing}{\spaceskip=\fontdimen2\font plus
\BIBentryALTinterwordstretchfactor\fontdimen3\font minus
  \fontdimen4\font\relax}
\providecommand{\BIBforeignlanguage}[2]{{%
\expandafter\ifx\csname l@#1\endcsname\relax
\typeout{** WARNING: IEEEtran.bst: No hyphenation pattern has been}%
\typeout{** loaded for the language `#1'. Using the pattern for}%
\typeout{** the default language instead.}%
\else
\language=\csname l@#1\endcsname
\fi
#2}}
\providecommand{\BIBdecl}{\relax}
\BIBdecl

\bibitem{fowler2011eradicating}
M.~Fowler, ``Eradicating non-determinism in tests,''
  \url{https://martinfowler.com/articles/nonDeterminism.html}, (Accessed on
  10/02/2024).

\bibitem{SandhuTesting2015}
\BIBentryALTinterwordspacing
A.~Sandhu, ``How to fix flaky tests,'' 2015. [Online]. Available:
  \url{https://tech.justeattakeaway.com/2015/03/30/how-to-fix-flaky-tests/}
\BIBentrySTDinterwordspacing

\bibitem{palmer2019}
\BIBentryALTinterwordspacing
J.~Palmer, ``Test flakiness – methods for identifying and dealing with flaky
  tests,'' 2019. [Online]. Available:
  \url{https://engineering.atspotify.com/2019/11/18/test-flakiness-methods-for-identifying-and-dealing-with-flaky-tests/}
\BIBentrySTDinterwordspacing

\bibitem{machalica2019predictive}
M.~Machalica, A.~Samylkin, M.~Porth, and S.~Chandra, ``Predictive test
  selection,'' in \emph{Proceedings of the International Conference on Software
  Engineering: Software Engineering in Practice (ICSE-SEIP)}, 2019.

\bibitem{lam2020dependent}
W.~Lam, A.~Shi, R.~Oei, S.~Zhang, M.~D. Ernst, and T.~Xie,
  ``Dependent-test-aware regression testing techniques,'' in \emph{Proceedings
  of the ACM International Symposium on Software Testing and Analysis (ISSTA)},
  2020.

\bibitem{thorve2018empirical}
S.~Thorve, C.~Sreshtha, and N.~Meng, ``An empirical study of flaky tests in
  android apps,'' in \emph{Proceedings of the IEEE International Conference on
  Software Maintenance and Evolution (ICSME)}, 2018.

\bibitem{googleFlaky2016}
J.~Micco, ``Google testing blog: Flaky tests at google and how we mitigate
  them,''
  \url{https://testing.googleblog.com/2016/05/flaky-tests-at-google-and-how-we.html},
  (Accessed on 25/03/2024).

\bibitem{Machalica2020Probabilistic}
M.~Machalica, W.~Chmiel, S.~Swierc, and R.~Sakevych, ``Probabilistic flakiness:
  How do you test your tests? - engineering at {M}eta,''
  \url{https://engineering.fb.com/2020/12/10/developer-tools/probabilistic-flakiness/},
  2020, (Accessed on 12/03/2024).

\bibitem{herzig2015art}
K.~Herzig, M.~Greiler, J.~Czerwonka, and B.~Murphy, ``The art of testing less
  without sacrificing quality,'' in \emph{Proceedings of the IEEE/ACM 37th IEEE
  International Conference on Software Engineering}, 2015.

\bibitem{Reducing45:online}
J.~Raine, ``Reducing flaky builds by 18x - the {GitHub} blog,''
  \url{https://github.blog/engineering/reducing-flaky-builds-by-18x/}, 2021,
  (Accessed on 31/03/2024).

\bibitem{luo2014empirical}
Q.~Luo, F.~Hariri, L.~Eloussi, and D.~Marinov, ``An empirical analysis of flaky
  tests,'' in \emph{Proceedings of the ACM International Symposium on
  Foundations of Software Engineering (FSE)}, 2014.

\bibitem{gruber2021empirical}
M.~Gruber, S.~Lukasczyk, F.~Kroi{\ss}, and G.~Fraser, ``An empirical study of
  flaky tests in python,'' in \emph{Proceedings of the IEEE Conference on
  Software Testing, Verification and Validation (ICST)}, 2021.

\bibitem{Hashemi2022flakyJS}
N.~Hashemi, A.~Tahir, and S.~Rasheed, ``An empirical study of flaky tests in
  javascript,'' in \emph{Proceedings of the IEEE International Conference on
  Software Maintenance and Evolution (ICSME)}, 2022.

\bibitem{eck2019understanding}
M.~Eck, F.~Palomba, M.~Castelluccio, and A.~Bacchelli, ``Understanding flaky
  tests: The developer's perspective,'' in \emph{Proceedings of the ACM Joint
  Meeting on European Software Engineering Conference and Symposium on the
  Foundations of Software Engineering (ESEC/FSE)}, 2019.

\bibitem{yost2023finding}
G.~A. Yost, ``Finding flaky tests in javascript applications using stress and
  test suite reordering,'' Master's thesis, The University of Texas at Austin,
  2023.

\bibitem{taleb2023frameworks}
M.~Taleb, ``{JavaScript} unit testing frameworks in 2024: A comparison ·
  {R}aygun blog,''
  \url{https://raygun.com/blog/javascript-unit-testing-frameworks/}, 2023,
  (Accessed on 09/02/2024).

\bibitem{jest_issue_4386}
J.~Community, ``Issue \#4386: Flakiness in tests with {J}est,''
  \url{https://github.com/jestjs/jest/issues/4386}, 2023, accessed: 2024-09-11.

\bibitem{zhang2014empirically}
S.~Zhang, D.~Jalali, J.~Wuttke, K.~Mu{\c{s}}lu, W.~Lam, M.~D. Ernst, and
  D.~Notkin, ``Empirically revisiting the test independence assumption,'' in
  \emph{Proceedings of the 2014 International Symposium on Software Testing and
  Analysis}, 2014, pp. 385--396.

\bibitem{Shi2019iFixFlakies}
A.~Shi, W.~Lam, R.~Oei, T.~Xie, and D.~Marinov, ``{iFixFlakies}: A framework
  for automatically fixing order-dependent flaky tests,'' in \emph{Proceedings
  of the ACM Joint Meeting on European Software Engineering Conference and
  Symposium on the Foundations of Software Engineering (ESEC/FSE)}, 2019.

\bibitem{reactTests}
``react-testing-library,''
  \url{https://github.com/testing-library/react-testing-library/blob/main/src/__tests__/auto-cleanup-skip.js},
  (Accessed on 10/02/2024).

\bibitem{stateofjs2023}
``State of {JavaScript} 2023: Testing,''
  \url{https://2023.stateofjs.com/en-US/libraries/testing/}, (Accessed on
  09/02/2024).

\bibitem{jestDescribeApi}
``Globals · {J}est,'' \url{https://jestjs.io/docs/api##describename-fn},
  (Accessed on 10/02/2024).

\bibitem{runningOrder}
``Option to run {J}est in band and sort tests alphabetically,''
  \url{https://github.com/jestjs/jest/issues/4032\#issuecomment-424427942},
  2018, (Accessed on 10/02/2024).

\bibitem{testSchedular}
``Jest test schedular,''
  \url{https://github.com/jestjs/jest/blob/1a487c1803124c594bdd86ee24b5949025660bc3/packages/jest-cli/src/test_scheduler.js\#L88-L96},
  (Accessed on 10/02/2024).

\bibitem{jestDocs}
``Setup and teardown · {J}est,''
  \url{https://jestjs.io/docs/setup-teardown#order-of-execution}, (Accessed on
  10/02/2024).

\bibitem{jestDocsCli}
``Jest cli options · jest,'' \url{https://jestjs.io/docs/cli}, (Accessed on
  10/02/2024).

\bibitem{testSequencer}
``Jest testsequencer,''
  \url{https://jestjs.io/docs/configuration#testsequencer-string}, (Accessed on
  10/02/2024).

\bibitem{Babel·Ba89:online}
``Babel · babel,'' \url{https://babel.dev/}, (Accessed on 10/01/2024).

\bibitem{RESTAPIe8:online}
``{REST} {API} endpoints for search - {GitHub} docs,''
  \url{https://docs.github.com/en/rest/search/search}, (Accessed on
  10/01/2024).

\bibitem{brisson2020we}
S.~Brisson, E.~Noei, and K.~Lyons, ``We are family: analyzing communication in
  {GitHub} software repositories and their forks,'' in \emph{Proceedings of the
  IEEE 27th International Conference on Software Analysis, Evolution and
  Reengineering (SANER)}.\hskip 1em plus 0.5em minus 0.4em\relax IEEE, 2020,
  pp. 59--69.

\bibitem{jestV20}
``Jest v20.0.0,'' \url{https://github.com/jestjs/jest/releases/tag/v20.0.0},
  (Accessed on 10/02/2024).

\bibitem{replicationPackage}
\BIBentryALTinterwordspacing
Anonymous, ``Detecting and evaluating order-dependent flaky tests in
  {JavaScript} - replication package,'' 2024. [Online]. Available:
  \url{https://doi.org/10.5281/zenodo.13852085}
\BIBentrySTDinterwordspacing

\bibitem{Jasmine-Matchers}
``Jasmine-matchers,'' \url{https://github.com/JamieMason/Jasmine-Matchers},
  (Accessed on 10/02/2024).

\bibitem{timkindb36:online}
``timkindberg/jest-when: Jest support for mock argument-matched return
  values.'' \url{https://github.com/timkindberg/jest-when}, (Accessed on
  10/01/2024).

\bibitem{tjcomman9:online}
``tj/commander.js: node.js command-line interfaces made easy,''
  \url{https://github.com/tj/commander.js}, (Accessed on 10/01/2024).

\bibitem{rbardini2:online}
``rbardini/jest-it-up: Automatically bump up global jest thresholds whenever
  coverage goes above them,'' \url{https://github.com/rbardini/jest-it-up},
  (Accessed on 10/01/2024).

\bibitem{fetch-mock-jest}
``fetch-mock-jest,''
  \url{https://github.com/wheresrhys/fetch-mock-jest/releases/tag/v1.5.1},
  (Accessed on 10/02/2024).

\bibitem{wang2022ipflakies}
R.~Wang, Y.~Chen, and W.~Lam, ``{iPFlakies}: A framework for detecting and
  fixing {P}ython order-dependent flaky tests,'' in \emph{Proceedings of the
  ACM/IEEE 44th International Conference on Software Engineering: Companion
  Proceedings}, 2022.

\bibitem{lam2019idflakies}
W.~Lam, R.~Oei, A.~Shi, D.~Marinov, and T.~Xie, ``{iDFlakies}: A framework for
  detecting and partially classifying flaky tests,'' in \emph{Proceedings of
  the IEEE Conference on Ssoftware Testing, Validation and Verification
  (ICST)}, 2019.

\bibitem{bell2014unit}
J.~Bell and G.~Kaiser, ``Unit test virtualization with {VMVM},'' in
  \emph{Proceedings of the 36th International Conference on Software
  Engineering}, 2014, pp. 550--561.

\bibitem{JestRunningTests:online}
``Building a {JavaScript} testing framework: Run all the tests in parallel,''
  \url{https://cpojer.net/posts/building-a-javascript-testing-framework#run-all-the-tests-in-parallel
  }, (Accessed on 10/01/2024).

\bibitem{gyori2015reliable}
A.~Gyori, A.~Shi, F.~Hariri, and D.~Marinov, ``Reliable testing: detecting
  state-polluting tests to prevent test dependency,'' in \emph{Proceedings of
  the International Symposium on Software Testing and Analysis (ISSTA)}, 2015.

\bibitem{gruber2023flapy}
M.~Gruber and G.~Fraser, ``Fla{P}y: mining flaky python tests at scale,'' in
  \emph{Proceedings of the IEEE/ACM International Conference on Software
  Engineering: Companion Proceedings (ICSE-Companion)}, 2023.

\bibitem{TheJestO19:online}
``The {J}est object · {J}est,''
  \url{https://jestjs.io/docs/jest-object#jestmockmodulename-factory-options},
  (Accessed on 10/01/2024).

\bibitem{amjed2022review}
A.~Tahir, S.~Rasheed, J.~Dietrich, N.~Hashemi, and L.~Zhang, ``Test
  flakiness’ causes, detection, impact and responses: A multivocal review,''
  \emph{Journal of Systems and Software}, vol. 206, p. 111837, 2023.

\bibitem{gambi2018practical}
A.~Gambi, J.~Bell, and A.~Zeller, ``Practical test dependency detection,'' in
  \emph{Proceedings of IEEE International Conference on Software Testing,
  Verification and Validation (ICST)}.\hskip 1em plus 0.5em minus 0.4em\relax
  IEEE, 2018.

\bibitem{rahman2024flakesync}
S.~Rahman and A.~Shi, ``{FlakeSync}: Automatically repairing async flaky
  tests,'' in \emph{Proceedings of the IEEE/ACM 46th International Conference
  on Software Engineering}, 2024.

\bibitem{rahman2024flakerake}
S.~Rahman, A.~Massey, W.~Lam, A.~Shi, and J.~Bell, ``Automatically reproducing
  timing-dependent flaky-test failures,'' in \emph{2024 IEEE Conference on
  Software Testing, Verification and Validation (ICST)}, 2024.

\bibitem{dong2020concurrency}
Z.~Dong, A.~Tiwari, X.~L. Yu, and A.~Roychoudhury, ``Flaky test detection in
  android via event order exploration,'' in \emph{Proceedings of the ACM Joint
  Meeting on European Software Engineering Conference and Symposium on the
  Foundations of Software Engineering (ESEC/FSE)}, 2021.

\bibitem{shi2016nondex}
A.~Shi, A.~Gyori, O.~Legunsen, and D.~Marinov, ``Detecting assumptions on
  deterministic implementations of non-deterministic specifications,'' in
  \emph{2016 IEEE International Conference on Software Testing, Verification
  and Validation (ICST)}, 2016.

\bibitem{dutta2020detecting}
S.~Dutta, A.~Shi, R.~Choudhary, Z.~Zhang, A.~Jain, and S.~Misailovic,
  ``Detecting flaky tests in probabilistic and machine learning applications,''
  in \emph{Proceedings of the ACM International Symposium on Software Testing
  and Analysis (ISSTA)}, 2020.

\bibitem{dutta2021fixing}
S.~Dutta, A.~Shi, and S.~Misailovic, ``{FLEX}: fixing flaky tests in machine
  learning projects by updating assertion bounds,'' in \emph{Proceedings of the
  29th ACM Joint Meeting on European Software Engineering Conference and
  Symposium on the Foundations of Software Engineering}, 2021.

\bibitem{rasheed2023effect}
S.~Rasheed, J.~Dietrich, and A.~Tahir, ``On the effect of instrumentation on
  test flakiness,'' in \emph{Proceedings of IEEE/ACM International Conference
  on Automation of Software Test (AST)}, 2023.

\bibitem{dietrich2022flaky}
J.~Dietrich, S.~Rasheed, and A.~Tahir, ``Flaky test sanitisation via on-the-fly
  assumption inference for tests with network dependencies,'' in
  \emph{Proceedings of the IEEE International Working Conference on Source Code
  Analysis and Manipulation (SCAM)}, 2022.

\bibitem{wei2021probabilistic}
A.~Wei, P.~Yi, T.~Xie, D.~Marinov, and W.~Lam, ``Probabilistic and systematic
  coverage of consecutive test-method pairs for detecting order-dependent flaky
  tests,'' in \emph{Tools and Algorithms for the Construction and Analysis of
  Systems: 27th International Conference (TACAS)}, 2021.

\bibitem{li2023systematically}
C.~Li, M.~M. Khosravi, W.~Lam, and A.~Shi, ``Systematically producing test
  orders to detect order-dependent flaky tests,'' in \emph{Proceedings of the
  32nd ACM SIGSOFT International Symposium on Software Testing and Analysis
  (ISSTA)}, 2023.

\bibitem{li2022repairing}
C.~Li, C.~Zhu, W.~Wang, and A.~Shi, ``Repairing order-dependent flaky tests via
  test generation,'' in \emph{Proceedings of the International Conference on
  Software Engineering (ICSE)}, 2022.

\bibitem{li2022evolution}
C.~Li and A.~Shi, ``Evolution-aware detection of order-dependent flaky tests,''
  in \emph{Proceedings of the ACM SIGSOFT International Symposium on Software
  Testing and Analysis (ISSTA)}, 2022.

\bibitem{parry2022evaluating}
O.~Parry, G.~M. Kapfhammer, M.~Hilton, and P.~McMinn, ``Evaluating features for
  machine learning detection of order-and non-order-dependent flaky tests,'' in
  \emph{Proceedings of the IEEE Conference on Software Testing, Verification
  and Validation (ICST)}, 2022.

\bibitem{costa2022test}
K.~Costa, R.~Ferreira, G.~Pinto, M.~d'Amorim, and B.~Miranda, ``Test flakiness
  across programming languages,'' \emph{IEEE Transactions on Software
  Engineering}, 2022.

\bibitem{fatima2024flakyFix}
S.~Fatima, H.~Hemmati, and L.~Briand, ``{FlakyFix}: Using large language models
  for predicting flaky test fix categories and test code repair,'' \emph{arXiv
  preprint arXiv:2307.00012}, 2024.

\bibitem{parry2021survey}
O.~Parry, G.~M. Kapfhammer, M.~Hilton, and P.~McMinn, ``A survey of flaky
  tests,'' \emph{ACM Transactions on Software Engineering and Methodology
  (TOSEM)}, vol.~31, no.~1, 2021.

\bibitem{zolfaghari2021root}
B.~Zolfaghari, R.~M. Parizi, G.~Srivastava, and Y.~Hailemariam, ``Root causing,
  detecting, and fixing flaky tests: State of the art and future roadmap,''
  \emph{Software: Practice and Experience}, vol.~51, no.~5, 2021.

\bibitem{king2018towards}
T.~M. King, D.~Santiago, J.~Phillips, and P.~J. Clarke, ``Towards a bayesian
  network model for predicting flaky automated tests,'' in \emph{Proceedings of
  the IEEE International Conference on Software Quality, Reliability and
  Security Companion (QRS-C)}, 2018.

\bibitem{endo2020noderacer}
A.~T. Endo and A.~M{\o}ller, ``{NodeRacer}: Event race detection for node.js
  applications,'' in \emph{Proceedings of the IEEE International Conference on
  Software Testing, Validation and Verification (ICST)}, 2020.

\end{thebibliography}

\end{document}